\documentclass[a4paper,10pt]{article}

\usepackage[dvips]{graphicx}
\usepackage{epsfig,amsmath,amssymb,verbatim,mathrsfs,array,layout,textcomp,amssymb,latexsym}

\newcommand{\beq}{\begin{eqnarray}}
\newcommand{\eeq}{\end{eqnarray}}

\def\beqa{\begin{eqnarray}}
\def\eeqa{\end{eqnarray}}
\newcommand{\no}{\nonumber}
\newcommand{\bv}{\left(\begin{array}{c}}
\newcommand{\ev}{\end{array}\right)}
\newcommand{\bmtwo}{\left(\begin{array}{cc}}
\newcommand{\bmthree}{\left(\begin{array}{ccc}}
\newcommand{\emn}{\end{array}\right)}
\newcommand{\bmtwoc}{\left\{\begin{array}{cc}}
\newcommand{\bmthreec}{\left\{\begin{array}{ccc}}
\newcommand{\emnc}{\end{array}\right\}}
\newcommand{\ba}{\begin{array}}
\newcommand{\ea}{\end{array}}

\def\lsim{\mathrel{\rlap{\lower4pt\hbox{\hskip1pt$\sim$}}
     \raise1pt\hbox{$<$}}}         
\def\gsim{\mathrel{\rlap{\lower4pt\hbox{\hskip1pt$\sim$}}
     \raise1pt\hbox{$>$}}}         

\begin{document}

\begin{titlepage}

\begin{flushright}
DO-TH 13/12
\end{flushright}

\vskip1.5cm
\begin{center}
  {\Large \bf Higgs couplings to fermions: 2HDM with MFV}
\end{center}
\vskip0.2cm

\begin{center}
Avital Dery$^1$, Aielet Efrati$^1$, Gudrun Hiller$^2$, Yonit Hochberg$^1$ and Yosef Nir$^1$\\
\end{center}
\vskip 8pt

\begin{center}
{\it $^1${}Department of Particle Physics and Astrophysics,
Weizmann Institute of Science, Rehovot 76100, Israel} \vspace*{0.2cm}
{\it $^2${}Institut f\"ur Physik, Technische Universit\"at Dortmund, D-44221 Dortmund, Germany}
{\tt  $^1${}avital.dery,aielet.efrati,yonit.hochberg,yosef.nir@weizmann.ac.il} \vspace*{0.2cm}
{\tt $^2${}ghiller@physik.uni-dortmund.de}
\end{center}

\vglue 0.3truecm

\begin{abstract}
  \vskip 3pt \noindent We clarify several subtleties concerning the implementation of minimal flavor violation (MFV) in two Higgs doublet models. We derive all the exact and approximate predictions of MFV for the neutral scalar ($h,H,A$) Yukawa couplings to fermions. We point out several possible tests of this framework at the LHC.
\end{abstract}

\end{titlepage}

\section{Introduction}
\label{sec:int}
New physics at or below the TeV scale, if coupled to the Standard Model~(SM) quarks and/or leptons, should have a very non-generic flavor structure. Otherwise, the contributions from the new physics to flavor changing neutral current~(FCNC) processes will be orders of magnitude above the experimental bounds. The question of how and why this non-generic structure arises is known as {\it the new physics flavor puzzle}. The most extreme solution to this puzzle is {\it minimal flavor violation~(MFV)}~\cite{D'Ambrosio:2002ex,Hall:1990ac,Chivukula:1987py,Buras:2000dm,Kagan:2009bn}.

In the absence of the Yukawa couplings, the SM gains a large non-Abelian global symmetry,
\beq
G_{\rm global}=SU(3)_Q\times SU(3)_U\times SU(3)_D\times SU(3)_L\times SU(3)_E.
\eeq
The three generations of a given gauge representation transform as a triplet under the corresponding $SU(3)$ factor. For example, the three left-handed quark doublets $Q_{1,2,3}$ transform as $(3,1,1,1,1)$ under $G_{\rm global}$. MFV is the idea that the only spurions that break the $[SU(3)]^5$ symmetry are the Yukawa matrices.

It is straightforward to implement MFV in new physics models with a single Higgs doublet or, more generally, in models with natural flavor conservation (NFC), where each sector -- $U$, $D$, $E$ -- is coupled to a single Higgs doublet. In all of these models, there are three, and only three, Yukawa matrices, in direct correspondence to the mass matrices:
\beq
Y^U=\frac{\sqrt2 M^U}{v_u},\ \ \ Y^D=\frac{\sqrt2 M^D}{v_d},\ \ \ Y^E=\frac{\sqrt2 M^E}{v_e},
\eeq
where $v_{u,d,e}$ is the vacuum expectation value of the Higgs doublet that couples to the up, down, and charged lepton sector, respectively. In case of a single Higgs, $v_u=v_d=v_e=v$. In this class of models, MFV is defined as follows~\cite{D'Ambrosio:2002ex}: All Lagrangian terms, constructed from SM fields and possibly new physics fields, and from the spurions
\beq
Y^U\sim(3,\bar3,1,1,1),\ \ \ Y^D\sim(3,1,\bar3,1,1),\ \ \ Y^E\sim(1,1,1,3,\bar3),
\eeq
are formally invariant under $G_{\rm global}$.

The question of how to implement MFV in multi-Higgs doublet models without NFC is more complicated. Think, for example, of a two Higgs doublet model~(2HDM). We denote the two doublets by $\Phi_1$ and $\Phi_2$. The Yukawa terms are given by
\beq\label{eq:lyuk}
{\cal L}_{\rm Yukawa}=-\sum_{i=1,2}\left(\overline Q\tilde\Phi_i Y^U_i U+\overline Q\Phi_i Y^D_i D+\overline L\Phi_i Y^E_i E+{\rm h.c.}\right).
\eeq
For each sector, $F=U,D,E$, there are two Yukawa matrices $Y^F_{1,2}$. In general, there is no reason to prefer one over the other as the basic spurion. Is there a loss of generality when we choose one or the other? Can we choose the mass matrices $(\sqrt2/v)M^F$ to play the role of spurions? What are the generic phenomenological features of 2HDMs with MFV regarding the Higgs-to-fermion couplings? In this work, we answer these questions.
Specifically, Section~\ref{sec:scalarcouplings} discusses scalar couplings, and Section~\ref{sec:spurion} addresses spurion choices. Predictions for the lepton (quark) sector are worked out in Section~\ref{sec:leptons} (Section~\ref{sec:quarks}). We discuss the phenomenological implications and summarize in Section~\ref{sec:discussion}.

\section{Scalar couplings in 2HDMs \label{sec:scalarcouplings}}
The Yukawa matrices in the Lagrangian~(\ref{eq:lyuk}) are responsible for the corresponding fermion mass matrices $M^F$, which we represent by the dimensionless matrices $Y^F_M$:
\beq
Y_M^F=\frac{\sqrt2 M^F}{v},\ \ \ (F=U,D,E).
\eeq
We denote by $Y_S^F, S=h,H,A$ the
 corresponding Yukawa couplings of the light CP-even scalar $h$, the heavy CP-even scalar $H$, and the CP-odd scalar $A$. (The Yukawa matrices of the charged Higgs $H^\pm$ are the same as those of $A$.) Each of these matrices is a linear combination of $Y_1$ and $Y_2$:
\beqa\label{eq:defys}
Y_M^F&=&+c_\beta Y_1^F+s_\beta Y_2^F,\nonumber\\
Y_A^F&=&-s_\beta Y_1^F+c_\beta Y_2^F,\nonumber\\
Y_h^F&=&-s_\alpha Y_1^F+c_\alpha Y_2^F,\nonumber\\
Y_H^F&=&+c_\alpha Y_1^F+s_\alpha Y_2^F,
\eeqa
where $c_\beta\equiv\cos\beta$, $s_\beta\equiv\sin\beta$, $\tan\beta\equiv v_2/v_1$, and $\alpha$ is the rotation angle from $({\cal R}e(\phi_1^0),{\cal R}e(\phi_2^0))$ to $(H,h)$. The angle $\beta$ is taken to be in the range $[0,\pi/2]$ while $\alpha\in[-\pi/2,+\pi/2]$.

We can express $Y_h$ and $Y_H$ in terms of $Y_M$ and $Y_A$:
\beqa\label{eq:hhma}
Y_h^F&=&c_{\alpha-\beta}Y_A^F-s_{\alpha-\beta}Y_M^F,\nonumber\\
Y_H^F&=&s_{\alpha-\beta}Y_A^F+c_{\alpha-\beta}Y_M^F.
\eeqa
We can also express $Y_M$ and $Y_A$ in terms of $Y_H$ and $Y_h$:
\beqa\label{eq:mhh}
Y_M^F&=&-s_{\alpha-\beta}Y_h^F+c_{\alpha-\beta}Y_H^F,\\
Y_A^F&=&+c_{\alpha-\beta}Y_h^F+s_{\alpha-\beta}Y_H^F.\label{eq:ahh}
\eeqa
In the corresponding mass basis, $Y_M^F$ is known:
\beqa\label{eq:ymmass}
U\ {\rm mass\ basis}:\ Y_M^U&=&\frac{\sqrt2}{v}{\rm diag}(m_u,m_c,m_t)=(y_u,y_c,y_t)\equiv\lambda_u,\nonumber\\
D\ {\rm mass\ basis}:\ Y_M^D&=&\frac{\sqrt2}{v}{\rm diag}(m_d,m_s,m_b)=(y_d,y_s,y_b)\equiv\lambda_d,\nonumber\\
E\ {\rm mass\ basis}:\ Y_M^E&=&\frac{\sqrt2}{v}{\rm diag}(m_e,m_\mu,m_\tau)=(y_e,y_\mu,y_\tau)\equiv\lambda_e.
\eeqa
In addition, $(\alpha-\beta)$ is known from the $hZZ$ coupling. This means that only one of the three Yukawa matrices $Y_{A,h,H}^F$ is independent: Measuring the entries of, say, $Y_h^F$ suffices to fix all entries of $Y_H^F$ (using (\ref{eq:mhh})) and then $Y_A^F$ (using~(\ref{eq:ahh})).

To demonstrate the power of (\ref{eq:mhh}), let us define a quantity that is, in principle, measurable:
\beq\label{eq:defxsmt}
W_{\tau\mu}^S\equiv\frac{(Y_S^E)_\tau}{y_\tau}-\frac{(Y_S^E)_\mu}{y_\mu}\,,\ \ \ S=h,H,A\,.
\eeq
Let us write Eq. (\ref{eq:mhh}) for two flavors separately:
\beqa
1&=&-s_{\alpha-\beta}\frac{(Y_h^E)_\tau}{y_\tau}+c_{\alpha-\beta}\frac{(Y_H^E)_\tau}{y_\tau},\nonumber\\
1&=&-s_{\alpha-\beta}\frac{(Y_h^E)_\mu}{y_\mu}+c_{\alpha-\beta}\frac{(Y_H^E)_\mu}{y_\mu}.
\eeqa
When we subtract these two equations, we obtain
\beq
\frac{W_{\tau\mu}^H}{W_{\tau\mu}^h}=\tan(\alpha-\beta)=\frac{c_V^h}{c_V^H}\,.
\eeq
Here $c_V^{h,H}$ is the strength of the coupling of the scalar to pairs of $Z$ or $W$ bosons relative to the SM. Similar relations hold for $W^{h,H}_{\mu e}$ and $W^{h,H}_{\tau e}$, and for the six pairs of quark flavors.

\section{Choosing the spurion \label{sec:spurion} }
As mentioned in the introduction, in the framework of a generic 2HDM, it is not obvious a-priori how to define the spurion. In this section we analyze this question. We focus here on the lepton sector, but our conclusion applies to the quark sector as well. For simplicity, in this section we suppress the super-index $E$ from the various $Y_Z^E$ matrices ($Z=1,2,M,S$) and MFV expansion coefficients introduced below wherever it is unambiguous.

In the absence of lepton Yukawa couplings, the global symmetry of the SM is
\beq
G_{\rm global}^\ell=SU(3)_L\times SU(3)_E.
\eeq
We define leptonic MFV as a situation where there is one, and only one, spurion that breaks $G_{\rm global}^\ell$:
\beq
\hat Y\sim(3,\bar3).
\eeq
(We leave the analysis of the case where neutrino-related spurions play a role to future work.)
In the most general case, each of the two leptonic Yukawa matrices is a power series in this spurion:
\beqa\label{eq:defabc}
Y_i&=&[a_i+b_i\hat Y\hat Y^\dagger+c_i(\hat Y\hat Y^\dagger)^2+\ldots ]\hat Y\,,\ \ \ i=1,2\,,
\eeqa
where the dots stand for terms of ${\cal O}(\hat Y^6)$ and higher. From here on we will not write the latter terms explicitly.

We define, for $x=a,b,c$,
\beqa
x_M&=&+c_\beta x_1+s_\beta x_2,\nonumber\\
x_A&=&-s_\beta x_1+c_\beta x_2,\nonumber\\
x_h&=&-s_\alpha x_1+c_\alpha x_2,\nonumber\\
x_H&=&+c_\alpha x_1+s_\alpha x_2.
\eeqa
Note that the relations (\ref{eq:hhma}) hold order by order in the MFV expansion:
\beqa\label{eq:xhham}
x_h&=&c_{\alpha-\beta}x_A-s_{\alpha-\beta}x_M,\nonumber\\
x_H&=&s_{\alpha-\beta}x_A+c_{\alpha-\beta}x_M.
\eeqa

We obtain,
\beqa\label{eq:abcm}
Y_M&=&[a_M +b_M\hat Y\hat Y^\dagger +c_M(\hat Y\hat Y^\dagger)^2]\hat Y,\\
Y_S&=&[a_S +b_S\hat Y\hat Y^\dagger +c_S(\hat Y\hat Y^\dagger)^2]\hat Y,\ \ \ S=A,h,H.\label{eq:abcs}
\eeqa
Using (\ref{eq:abcm}), we can write the spurion $\hat Y$ as a power series in $Y_M$:
\beq\label{eq:hatyym}
\hat Y=\left[1-\frac{b_M}{a_M^3}Y_M Y_M^\dagger+\frac{(3b_M^2-a_M c_M)}{a_M^6}(Y_M Y_M^\dagger)^2\right]\frac{Y_M}{a_M}.
\eeq
We obtain, for $S=A,h,H$:
\beq\label{eq:ysm}
Y_S&=&\left\{\frac{a_S}{a_M}+\left(b_S-b_M \frac{a_S}{a_M}\right)\frac{Y_MY_M^\dagger}{a_M^3}+\left[c_S-c_M\frac{a_S}{a_M}-\frac{3b_M}{a_M}\left(b_S-b_M \frac{a_S}{a_M}\right)\right]\frac{(Y_MY_M^\dagger)^2}{a_M^5}\right\}Y_M\no\\
&\equiv&\left[A_S+B_S Y_MY_M^\dagger+C_S(Y_MY_M^\dagger)^2\right]Y_M\,.\label{eq:ys}
\eeq

Eqs. (\ref{eq:hatyym}) and (\ref{eq:ys}) lead to an important conclusion: If $a_M\neq0$, then one can use $Y_M$ as the spurion. Indeed, if there is a `fundamental' spurion $\hat Y$ that transforms as $(3,\bar3)$ under $SU(3)_L\times SU(3)_E$, then any matrix $Y_Z$ that breaks this symmetry as $(3,\bar3)$ and, in the expansion in $\hat Y$, has a term linear in $\hat Y$, namely $a_Z\neq0$, can be used as a spurion without loss of generality. This matrix can be $Y_1$, if $a_1\neq0$, or $Y_2$, if $a_2\neq0$, or any linear combination of them, such as $Y_M$ and $Y_S$. While we do not really know whether any of $a_i,a_M,a_S$ vanishes, there is no reason to expect that this would be the case.

Note that all the matrices $Y_Z$ that have $a_Z\neq0$ are proportional to each other at leading order in the spurion, but differ at the next to leading order. For example, we have
\beq
\frac{(Y_M)_\mu}{(Y_M)_\tau}=\frac{\hat Y_\mu}{\hat Y_\tau}\left[1+(b_M/a_M)(\hat Y_\mu^2-\hat Y_\tau^2)+{\cal O}(\hat Y^4)\right].
\eeq
One may wonder then how an expansion in different spurions, such as $\hat Y\to Y_M$,  would yield the same predictions. The answer is that expansions in different spurions involve also different higher order corrections, $b_S\to B_S$, {\it etc.}, such that the results remain the same. Since anyway neither $x_S$ nor $X_S$ are known, there is no loss of generality.

Thus, we can choose any matrix $Y_Z$ with $a_Z\neq0$ as our spurion, without loss of generality. Since $Y_M$ is already measured and known -- see Eq.~(\ref{eq:ymmass}) -- it is a particularly convenient choice for a spurion in generic MFV 2HDMs. From here on we use Eq.~(\ref{eq:ys}) as the starting point for obtaining the MFV predictions.

There is, however, one point that is worth clarifying. While all spurions $Y_Z$  with  $a_Z\neq0$ are proportional to each other at leading order, their overall size can be very different from each other. Thus, while all entries of the leptonic $Y_M^E$ are small, this is not necessarily the case for the `fundamental' spurion $\hat Y^E$ and, consequently, for $Y_S^E$. Specifically, we expect $a_M^E$ to be in the range $m_\tau/v\lsim a_M^E\lsim1$. In case that $a_M^E\sim1$, higher order terms in the spurion expansion are always small, and relations obtained at a given order remain approximately correct to all orders. However, in case that $a_M^E\sim m_\tau/v$, terms that are higher order in $\hat Y_\tau$ might modify the lower order relations in a significant way.

In case that $\hat Y_\tau={\cal O}(1)$, we expect that some of the $(Y_S)_\tau$ are also of order one. Given the following sum rule, which follows from Eq. (\ref{eq:defys}),
\beq
(Y_M)_\tau^2+(Y_A)_\tau^2=(Y_h)_\tau^2+(Y_H)_\tau^2,
\eeq
and the fact that $(Y_M)_\tau\ll1$, it is clear that if any $(Y_S)_\tau={\cal O}(1)$, then at least two of the three are ${\cal O}(1)$ and $(Y_A)_\tau$ must be one of them. For these, we will have $A_S^E\sim1/y_\tau$, $B_S^E\lsim 1/y_\tau^3$, $C_S^E\lsim1/y_\tau^5$, {\it etc.}, compensating for $y_\tau\ll\hat Y_\tau$. In this sense, the expansion in $Y_M^E$ might be misleading: It gives the naive impression that terms that are higher order in $y_\tau^2$ are small, while if $\hat Y_\tau\sim1$, we may have $1\ll A_S^E \ll B_S^E\ll C_S^E\ll \ldots$, possibly making the higher order terms comparable to the lower order ones. In what follows, we use `$(Y_A)_\tau\sim1$' to define the case where higher order corrections might invalidate relations obtained at a certain order.

The situation in the MFV models where $1/a_M^E\gg1$ is similar to the situation in the NFC models where $\tan\beta\gg1$. Indeed, from Eq. (\ref{eq:defys}), we learn that we can have $(Y_M)_\tau\ll1$ and $(Y_A)_\tau={\cal O}(1)$ only if $(Y_2)_\tau\ll1$ and $c_\beta\ll1$, namely large $\tan\beta$. (The other possibility, $(Y_1)_\tau\ll1$ and $s_\beta\ll1$ is just a change of notations, $1\leftrightarrow2$.)

\section{Predictions of minimal lepton flavor violation \label{sec:leptons}}
In this section we continue to omit the super-index $E$ from the various $Y_Z^E$ matrices.

It is clear from Eq.~(\ref{eq:ys}) that all of $Y_h$, $Y_H$ and $Y_A$ are diagonal in the charged lepton mass basis. We learn that minimal lepton flavor violation (MLFV) (with a single spurion that breaks $SU(3)_L$) predicts
\beq\label{eq:ysab}
(Y_S)_{\ell \ell'}=0\ {\rm for\ flavors}\ \ell\neq\ell'.
\eeq

The nine diagonal couplings $(Y_S)_\ell\equiv(Y_S)_{\ell\ell}$, can all, in principle, be measured. Eqs.~(\ref{eq:mhh}) and~(\ref{eq:ahh}) imply, however, that for each lepton flavor $\ell$, only one of the three couplings $(Y_{h,H,A})_\ell$ is independent. We thus have three observables to test further MLFV predictions.

In addition to the exact MLFV predictions of Eq. (\ref{eq:ysab}), one can obtain several approximate predictions. Let us start by writing the explicit expressions for $(Y_S)_{e,\mu,\tau}$ to ${\cal O}(y_\ell^5)$:
\beqa
(Y_S)_\tau&=&(A_S+B_S y_\tau^2+C_S y_\tau^4)y_\tau,\nonumber\\
(Y_S)_\mu&=&(A_S+B_S y_\mu^2+C_S y_\mu^4)y_\mu,\nonumber\\
(Y_S)_e&=&(A_S+B_S y_e^2+C_S y_e^4)y_e.
\eeqa

First, we take into account only the ${\cal O}(Y_M)$ terms, namely take
\beq
B_S=C_S=0.
\eeq
The three couplings then depend on a single parameter, $A_S$. Thus, if we know one coupling, we can predict the other two. The approximate MLFV relations read:
\beqa\label{eq:univmt}
\frac{(Y_S)_\mu}{(Y_S)_\tau}&=&\frac{m_\mu}{m_\tau},\\
\frac{(Y_S)_e}{(Y_S)_\mu}&=&\frac{m_e}{m_\mu}.\label{eq:univem}
\eeqa
These relations we call {\it universality}. Note that the relation (\ref{eq:univmt}) can be significantly corrected if $(Y_A)_\tau\sim1$. As concerns (\ref{eq:univem}), the higher order corrections are always small: Even for  $(Y_A)_\tau\sim1$, the violation is of order $(y_\mu/y_\tau)^2$.

Second, we take into account also the ${\cal O}(Y_M^3)$ terms, but still neglect the higher order terms, namely take
\beq
C_S=0.
\eeq
The three couplings depend on two parameters, $A_S$ and $B_S$. Thus, if we know two couplings, we can predict the third. The approximate relation among the non-universal effects can be deduced by using
\beq\label{eq:abmt}
\frac{(Y_S)_\mu^2}{(Y_S)_\tau^2}=\frac{m_\mu^2}{m_\tau^2}
\left[1+4\frac{B_S}{A_S}\frac{(m_\mu^2-m_\tau^2)}{v^2}\right],
\eeq
and noticing that $A_S$ and $B_S$ are generation independent. We obtain
\beq\label{eq:nonuniv}
\frac{\left[(Y_S)_e/(Y_S)_\tau\right]^2-\left(m_e/m_\tau\right)^2}
{\left[(Y_S)_\mu/(Y_S)_\tau\right]^2-\left(m_\mu/m_\tau\right)^2}=
\frac{m_e^2}{m_\mu^2}\left(1+\frac{m_\mu^2-m_e^2}{m_\tau^2}\right).
\eeq

When we add the $C_S$ terms, we have three couplings which depend on three parameters, so there are no parameter-independent relations anymore. Yet, Eq.~(\ref{eq:nonuniv}) receives no corrections of order $y_\tau^2$ and thus remains a good approximation even if $(Y_A)_\tau={\cal O}(1)$. To see this, we write
\beq\label{eq:abcmt}
\frac{(Y_S)_\mu^2}{(Y_S)_\tau^2}-\frac{m_\mu^2}{m_\tau^2}=
\frac{m_\mu^2}{m_\tau^2}\frac{4B_S}{A_S}\frac{(m_\mu^2-m_\tau^2)}{v^2}\left[1+
\frac{(B_S^2+2A_SC_S)}{A_S B_S}\frac{(m_\mu^2+m_\tau^2)}{v^2}
-\frac{4B_S}{A_S}\frac{m_\tau^2}{v^2}\right],
\eeq
and similarly for $(Y_S)_e^2/(Y_S)_\tau^2-m_e^2/m_\tau^2$. Then, for the ratio, we obtain
\beq
\frac{[(Y_S)_e/(Y_S)_\tau]^2-(m_e/m_\tau)^2}{[(Y_S)_\mu/(Y_S)_\tau]^2-(m_\mu/m_\tau)^2}=
\frac{m_e^2}{m_\mu^2}\left[1+\frac{m_\mu^2-m_e^2}{m_\tau^2}
+\frac{B_S^2+2A_SC_S}{A_SB_S}\times\frac{m_e^2-m_\mu^2}{v^2}\right].
\eeqa
Thus, indeed, the leading corrections to Eq. (\ref{eq:nonuniv}) are of ${\cal O}(y_\mu^2/y_\tau^2)$ and therefore small.

In the framework of 2HDM, a class of constrained MFV models, where (see Eq. (\ref{eq:defabc}))
\beq
b_{1,2}=c_{1,2}=\ldots=0,
\eeq
is known as `alignment models' \cite{Pich:2009sp}. These models give universality, Eqs. (\ref{eq:univmt},\ref{eq:univem}) as an exact relation, rather than an approximate one as in general MLFV. As concerns Eq. (\ref{eq:nonuniv}), both the numerator and the denominator of this equation are zero in alignment models, and thus the equation is not well defined.

To summarize,
MLFV predicts that the leptonic off-diagonal Yukawa couplings vanish. As concerns the diagonal couplings, there are three independent ones. Their overall size is never fixed by MLFV. The following approximate relations are, however, predicted:
\begin{itemize}
\item Knowing $(Y_S)_\tau$ allows us to fix $(Y_S)_{\mu,e}$ up to corrections of order $\hat Y_\tau^2$.
\item Knowing $(Y_S)_\mu$ allows us to  fix $(Y_S)_e$ up to corrections of order $\hat Y_\mu^2$.
\item Knowing $(Y_S)_\mu/(Y_S)_\tau$ allows us to fix  $(Y_S)_e/(Y_S)_\tau$ up to corrections of order $\hat Y_\mu^2$.
\item  Knowing $(Y_S)_e/(Y_S)_\mu$ allows us to fix  $(Y_S)_\mu/(Y_S)_\tau$ up to corrections of order $\hat Y_\tau^2$.
\end{itemize}
If $(Y_A)_\tau\sim1$, then the predictions corrected by powers of $\hat Y_\tau$ (first and fourth item) are not reliable.

\section{Predictions of minimal quark flavor violation \label{sec:quarks}}

Here we work out predictions for 2HDMs with minimal quark flavor violation. We discuss the spurions in
Section~\ref{sec:quarkspurions} and the implications for the up and down sectors in Sections~\ref{sec:up} and~\ref{sec:down}, respectively.  We summarize the quark MFV predictions in Section
\ref{sec:quarkpredictions}.

\subsection{Quark flavor spurions \label{sec:quarkspurions}}
In the absence of quark Yukawa couplings, the global symmetry is
\beq
G_{\rm global}^q=SU(3)_Q\times SU(3)_U\times SU(3)_D.
\eeq
The basic idea of MFV is that there are two, and only two, spurions that break $G_{\rm global}^q$:
\beq
\hat Y^U(3,\bar3,1),\ \ \ \hat Y^D(3,1,\bar3).
\eeq
In the most general case, each of the four Yukawa matrices is a power series in these spurions ($i=1,2$):
\beqa
Y^U_i&=&[a^U_i+b^U_i\hat Y^U\hat Y^{U\dagger}+c^U_i\hat Y^D\hat Y^{D\dagger}+\ldots]\hat Y^U,\nonumber\\
Y^D_i&=&[a^D_i+b^D_i\hat Y^D\hat Y^{D\dagger}+c^D_i\hat Y^U\hat Y^{U\dagger}+\ldots]\hat Y^D,
\eeqa
where the dots stand for terms of ${\cal O}(\hat Y^4)$ and higher,  which we suppress in the following.
The quark mass matrices are similarly power series in the spurions:
\beqa
Y^U_M\equiv\frac{\sqrt2 M^U}{v}&=&\left[a^U_M+b^U_M\hat Y^U\hat Y^{U\dagger}+c^U_M\hat Y^D\hat Y^{D\dagger}\right]\hat Y^U, \nonumber\\
Y^D_M\equiv\frac{\sqrt2 M^D}{v}&=&\left[a^D_M+b^D_M\hat Y^D\hat Y^{D\dagger}+c^D_M\hat Y^U\hat Y^{U\dagger}\right]\hat Y^D.
\eeqa
Assuming that $a^{U,D}_M\neq0$, we can then write the spurions $\hat Y^{U,D}$ as power series in $Y^{U,D}_M$:
\beqa
\hat Y^U&=&\left[1-\frac{b^U_M}{a^{U3}_M}Y^U_M Y^{U\dagger}_M-\frac{c^U_M}{a^{D2}_M a^U_M}Y^D_M Y^{D\dagger}_M\right]\frac{Y^U_M}{a^U_M}\,,\nonumber\\
\hat Y^D&=&\left[1-\frac{b^D_M}{a^{D3}_M}Y^D_M Y^{D\dagger}_M-\frac{c^D_M}{a^{U2}_M a^D_M}Y^U_M Y^{U\dagger}_M\right]\frac{Y^D_M}{a^D_M}\,.
\eeqa
Each of $S=h,H,A$ has Yukawa matrices to the up and down sectors, $Y^{U,D}_S$.
As argued in Section \ref{sec:scalarcouplings}, however, only one of the three $Y_S^U$ is independent, and similarly only one of the three $Y_S^D$.

The same arguments that we presented in discussing MLFV imply that we can use $Y^{U,D}_M$ as our spurions:
\beqa
Y_S^U&=&\left(A_S^U+B_S^U Y_M^UY_M^{U\dagger}+C_S^U Y_M^D Y_M^{D\dagger}\right)Y_M^U,\nonumber\\
Y_S^D&=&\left(A_S^D+B_S^D Y_M^DY_M^{D\dagger}+C_S^D Y_M^U Y_M^{U\dagger}\right)Y_M^D,
\eeqa
where
\beqa
A_S^U&=&\frac{a_S^U}{a_M^U},\ \ \
B_S^U=\frac{b_S^U-b_M^U a_S^U/a_M^U}{a_M^{U3}},\ \ \
C_S^U=\frac{c_S^U-c_M^U a_S^U/a_M^U}{a_M^U a_M^{D2}}\nonumber , \\
A_S^D&=&\frac{a_S^D}{a_M^D},\ \ \
B_S^D=\frac{b_S^D-b_M^D a_S^D/a_M^D}{a_M^{D3}},\ \ \
C_S^D=\frac{c_S^D-c_M^D a_S^D/a_M^D}{a_M^D a_M^{U2}}.
\eeqa

In the respective mass bases, we have:
\beqa\label{eq:ysumass}
U\ {\rm mass\ basis}:\
Y^U_S&=&(A^U_S+B^U_S\lambda_u^2+C^U_S V\lambda_d^2 V^\dagger)\lambda_u,\\
D\ {\rm mass\ basis}:\
Y^D_S&=&(A^D_S+B^D_S\lambda_d^2+C^D_S V^\dagger\lambda_u^2 V)\lambda_d,\label{eq:ysdmass}
\eeqa
where $V$ is the CKM matrix.

We again caution the reader that the small size of all entries in $\lambda_d$ (and in particular $y_b\ll1$) might be misleading. If $(Y_A^D)_b\sim1$, then we expect $A_S^D\sim1/y_b$, and may have $B_S^D\sim1/y_b^3$, $C_S^D\sim1/y_b$ and $C_S^U\sim1/y_b^2$ for $Y_A^D$ and at least one of $Y_h^D$ and $Y_H^D$. Since $y_t\sim1$, higher order terms in the MFV expansion involving $y_t$ can induce ${\cal O}(1)$ effects and resummation is needed~\cite{Kagan:2009bn}.

\subsection{The up sector \label{sec:up}}
The ratios between the off-diagonal terms in the up sector are fixed by CKM and mass ratios. The largest off-diagonal term is $(Y^U_S)_{ct}$. The largest diagonal term is $(Y^U_S)_{tt}$. The ratio between the two is given by
\beq
\frac{(Y^U_S)_{ct}}{(Y^U_S)_{tt}}=\frac{C^U_S y_b^2}{A^U_S+B^U_S y_t^2+C^U_Sy_b^2|V_{tb}|^2}\ V_{cb}V_{tb}^*,
\eeq
which, for $(Y_A^D)_b\sim1$, could be as large as $V_{cb}$.

As concerns the ratios among the off-diagonal couplings, there are two relations that remain good approximations to all orders:
\beqa
\frac{(Y^U_S)_{ut}}{(Y^U_S)_{ct}}&=&\frac{V_{ub}}{V_{cb}},\nonumber\\
\frac{(Y^U_S)_{tu}}{(Y^U_S)_{tc}}&=&\frac{V_{ub}^*}{V_{cb}^*}\frac{m_u}{m_c}.
\eeqa
The other three independent ratios suffer from ${\cal O}(1)$ corrections:
\beqa
\frac{(Y^U_S)_{cu}}{(Y^U_S)_{uc}}&=&\frac{V_{ub}^*V_{cb}}{V_{ub}V_{cb}^*}\
\frac{\left(1+\frac{m_s^2}{m_b^2}\frac{V_{cs}V_{us}^*}{V_{cb}V_{ub}^*}\right)}
{\left(1+\frac{m_s^2}{m_b^2}\frac{V_{us}V_{cs}^*}{V_{ub}V_{cb}^*}\right)}\ \frac{m_u}{m_c},\nonumber\\
\frac{(Y^U_S)_{tc}}{(Y^U_S)_{ct}}&=&\frac{V_{tb}V_{cb}^*}{V_{cb}V_{tb}^*}\ \frac{m_c}{m_t},\nonumber\\
\frac{(Y^U_S)_{uc}}{(Y^U_S)_{ct}}&=&\frac{V_{ub}V_{cb}^*}{V_{cb}V_{tb}^*}
\left(1+\frac{m_s^2}{m_b^2}\frac{V_{us}V_{cs}^*}{V_{ub}V_{cb}^*}\right)\frac{m_c}{m_t}.
\eeqa
In the above, the first ratio suffers from ${\cal O}(1)$ corrections if $(Y_A^D)_b\sim 1$, but remains a good approximation if $(Y_A^D)_b\ll1$, while the last two ratios suffer from ${\cal O}(1)$ corrections regardless of the size of $(Y_A^D)_b$.

As concerns the diagonal terms, we can obtain several approximate predictions. Let us start by writing the explicit expressions for $(Y_S^U)_{u,c,t}$ to ${\cal O}(y_q^3)$:
\beqa
(Y_S^U)_t&=&\left[A_S^U+B_S^U y_t^2+C_S^U y_b^2|V_{tb}|^2\right]y_t,\nonumber\\
(Y_S^U)_c&=&\left[A_S^U+B_S^U y_c^2+C_S^U
\left(y_b^2|V_{cb}|^2+y_s^2|V_{cs}|^2\right)\right]y_c,\nonumber\\
(Y_S^U)_u&=&\left[A_S^U+B_S^U y_u^2+C_S^U
\left(y_b^2|V_{ub}|^2+y_s^2|V_{us}|^2\right)\right]y_u.
\eeqa
First, we take into account only the ${\cal O}(Y_M)$ terms, namely take
\beq
B_S^U=C_S^U=0.
\eeq
The three couplings then depend on a single parameter $A_S^U$. Thus, if we know one coupling, we can predict the other two. The approximate MFV relations read:
\beqa\label{eq:uc}
\frac{(Y^U_S)_u}{(Y^U_S)_c}&=&\frac{m_u}{m_c},\\
\frac{(Y^U_S)_c}{(Y^U_S)_t}&=&\frac{m_c}{m_t}.\label{eq:ct}
\eeqa
In general, however, the relation~(\ref{eq:ct}) is significantly corrected by ${\cal O}(y_t^2)$ terms. As concerns (\ref{eq:uc}), the higher order terms are always small: Even for $(Y_A^D)_b\sim1$, the violation is of order $|V_{cb}|^2$.

Second, we argue that there is one relation between the violations of universality that remains an excellent approximation even when higher order terms are taken into account, and that is
\beq\label{eq:ucnu}
\frac{(Y^U_S)_u/(Y^U_S)_t-m_u/m_t}{(Y^U_S)_c/(Y^U_S)_t-m_c/m_t}
=\frac{m_u}{m_c}.
\eeq

In the case that $\hat Y_b\sim y_b$, we can neglect the ${\cal O}(Y^{D2}_MY^U_M)$ terms, namely take
\beq
C^U_S=0.
\eeq
Then the three couplings depend on two parameters, $A_S^U$ and $B_S^U$. Thus, if we know two couplings, we can predict the third. We obtain
\beq\label{eq:ucb}
\frac{(Y^U_S)_u/(Y^U_S)_t-m_u/m_t}{(Y^U_S)_c/(Y^U_S)_t-m_c/m_t}=
\frac{m_u}{m_c}\left(1+\frac{m_c^2-m_u^2}{m_t^2}\right).
\eeq
Thus the correction to (\ref{eq:ucnu}) is tiny, of order $m_c^2/m_t^2\sim10^{-4}$.

If $(Y_A)_b\sim1$, we can neglect neither $C_S^U$ nor the $B_S^U$. Consequently, there is no relation left which is independent of the unknown expansion parameters. We obtain:
\beq
\frac{(Y^U_S)_u/(Y^U_S)_t-m_u/m_t}{(Y^U_S)_c/(Y^U_S)_t-m_c/m_t}=
\frac{m_u}{m_c}
\left\{1+\frac{B^U_S(y_c^2-y_u^2)
+C^U_S[y_b^2(|V_{cb}|^2-|V_{ub}|^2)+y_s^2(|V_{cs}|^2-|V_{us}|^2)]}
{B^U_S y_t^2+C^U_Sy_b^2|V_{tb}|^2}\right\}.
\eeq
Thus the correction to (\ref{eq:ucnu}) is at most of ${\cal O}(|V_{cb}|^2)\sim10^{-3}$.

\subsection{The down sector \label{sec:down}}
The ratios between the off-diagonal terms in the down sector are fixed by CKM and mass ratios. The largest off-diagonal term is $(Y^D_S)_{sb}$. The largest diagonal term is $(Y_S^D)_{bb}$. The ratio between the two is given by
\beq
\frac{(Y^D_S)_{sb}}{(Y^D_S)_{bb}}=\frac{C_S^D y_t^2}{A_S^D +B_S^Dy_b^2+C_S^Dy_t^2|V_{tb}|^2}V_{ts}^*V_{tb}.
\eeq
If $C_S^D\not\ll A_S^D$, then this ratio is of order  $V_{ts}$.

As concerns the ratios among the off-diagonal couplings, there are three relations that remain good approximations to all orders, even for $(Y_A^D)_b\sim1$:
\beqa
\frac{(Y^D_S)_{db}}{(Y^D_S)_{sb}}&=&\frac{V_{td}^*}{V_{ts}^*},\nonumber\\
\frac{(Y^D_S)_{bd}}{(Y^D_S)_{bs}}&=&\frac{V_{td}}{V_{ts}}\frac{m_d}{m_s},\nonumber\\
\frac{(Y^D_S)_{sd}}{(Y^D_S)_{ds}}&=&\frac{V_{td}V_{ts}^*}{V_{td}^*V_{ts}}\frac{m_d}{m_s}.
\eeqa
The other two independent ratios suffer from ${\cal O}(1)$ corrections if $(Y_A^D)_b\sim1$, and remain good approximations if $(Y_A^D)_b\ll1$:
\beqa
\frac{(Y^D_S)_{bs}}{(Y^D_S)_{sb}}&=&\frac{V_{tb}^*V_{ts}}{V_{ts}^*V_{tb}}\ \frac{m_s}{m_b},\nonumber\\
\frac{(Y^D_S)_{ds}}{(Y^D_S)_{sb}}&=&\frac{V_{td}^*V_{ts}}{V_{ts}^*V_{tb}}\ \frac{m_s}{m_b}.
\eeqa

As concerns the diagonal terms, we can obtain several approximate predictions. We start by writing the explicit expressions for $(Y_S^D)_{d,s,b}$ to ${\cal O}(y_q^3)$:
\beqa
(Y_S^D)_b&=&\left[A_S^D+B_S^D y_b^2+C_S^D y_t^2|V_{tb}|^2\right]y_b,\nonumber\\
(Y_S^D)_s&=&\left[A_S^D+B_S^D y_s^2+C_S^D y_t^2|V_{ts}|^2\right]y_s,\nonumber\\
(Y_S^D)_d&=&\left[A_S^D+B_S^D y_d^2+C_S^D
y_t^2|V_{td}|^2\right]y_d.
\eeqa
First, we take into account only the ${\cal O}(Y_M)$ terms, namely take
\beq
B_S^D=C_S^D=0.
\eeq
The three couplings then depend on a single parameter $A_S^U$. Thus, if we know one coupling, we can predict the other two. The approximate MFV relations read:
\beqa\label{eq:ds}
\frac{(Y^D_S)_d}{(Y^U_S)_s}&=&\frac{m_d}{m_s},\\
\frac{(Y^D_S)_s}{(Y^U_S)_b}&=&\frac{m_s}{m_b}.\label{eq:sb}
\eeqa
In general, however, the relation (\ref{eq:sb}) suffers from large, ${\cal O}(y_t^2|V_{tb}|^2)$ corrections.

Second, we argue that there is one relation between the violations of universality that remains an excellent approximation even when higher order terms are taken into account, and that is
\beq\label{eq:dsnu}
\frac{(Y^D_S)_d/(Y^D_S)_b-m_d/m_b}{(Y^D_S)_s/(Y^D_S)_b-m_s/m_b}
=\frac{m_d}{m_s}.
\eeq
In the case that $\hat Y_b\sim y_b$, we can take into account the ${\cal O}(Y_M^{U2}Y_M^D)$ terms, but still neglect the ${\cal O}(Y_M^{D3})$ terms, namely take
\beq
B_S^D=0.
\eeq
The three couplings depend on two parameters, $A_S^D$ and $C_S^D$, implying a single relation between the couplings that is independent of the expansion coefficients:
\beq\label{eq:ysdyss}
\frac{(Y^D_S)_d/(Y^D_S)_b-m_d/m_b}{(Y^D_S)_s/(Y^D_S)_b-m_s/m_b}=
\frac{m_d}{m_s}
\left(1+\frac{|V_{ts}|^2-|V_{td}|^2}{|V_{tb}|^2}\right).
\eeq
Thus, the correction to (\ref{eq:dsnu}) is of order $|V_{ts}/V_{tb}|^2\sim10^{-3}$.

When we add the $B_S^D$ terms, there are no relations that are independent of unknown expansion coefficients. We obtain:
\beq\label{eq:ysdyssb}
\frac{(Y^D_S)_d/(Y^D_S)_b-m_d/m_b}{(Y^D_S)_s/(Y^D_S)_b-m_s/m_b}=
\frac{m_d}{m_s}
\left[1+\frac{B^D_S(y_s^2-y_d^2)
+C^D_S y_t^2(|V_{ts}|^2-|V_{td}|^2)}
{B^D_S y_b^2+C^D_S y_t^2|V_{tb}|^2}\right].
\eeq
Thus the corrections to (\ref{eq:dsnu}) are of order $10^{-3}$.

\subsection{Quark MFV relations \label{sec:quarkpredictions}}

To summarize, MFV in the quark sector predicts a large number of approximate relations:
\begin{itemize}
\item Knowing one off-diagonal entry in $Y_S^U$ allows us to fix the corresponding $u\leftrightarrow c$ off-diagonal entry if $(Y_A^D)_b\ll 1$, and all other off-diagonal entries are then known up to ${\cal O}(1)$ corrections. The same is true if $(Y_A^D)_b\sim 1$, except that $(Y_S^U)_{cu}$ and $(Y_S^U)_{uc}$ are then related to each other up to ${\cal O}(1)$ corrections only.
\item Knowing one off-diagonal entry in $Y_S^D$ allows us to fix all off-diagonal entries if $(Y_A^D)_b\ll1$. If $(Y_A^D)_b\sim1$, knowing one off-diagonal entry in $Y_S^D$ allows us to fix the corresponding $d\leftrightarrow s$ off-diagonal entry, and all other off-diagonal entries are then known up to ${\cal O}(1)$ corrections.
\item Knowing $(Y^U_S)_c$ allows us to fix $(Y^U_S)_u$ up to corrections of order $\hat Y_c^2$.
\item Knowing $(Y^D_S)_s$ allows us to fix $(Y^D_S)_d$ up to corrections of order $\hat Y_s^2$.
\item Knowing $(Y^U_S)_c/(Y^U_S)_t$ allows us to fix $(Y^U_S)_u/(Y^U_S)_t$ up to corrections of order $(\hat Y_b/\hat Y_t)^2|V_{cb}|^2$.
\item Knowing $(Y^D_S)_s/(Y^D_S)_b$ allows us to fix $(Y^D_S)_d/(Y^D_S)_b$ up to corrections of order $|V_{ts}|^2$.
\end{itemize}
As argued in Section~\ref{sec:scalarcouplings}, no additional flavor information can be obtained from measurements of scalar couplings beyond $Y^{U,D}_h$, namely those of $S=H, A$.

\section{Discussion \& Summary  \label{sec:discussion}}
Minimal flavor violation (MFV) is the assumption that there are two, and only two, spurions that break the global $SU(3)_Q\times SU(3)_U\times SU(3)_D$ flavor symmetry, with one of them transforming as $(3,\bar3,1)$ and the other as $(3,1,\bar3)$ under this symmetry. In this work, our definition of minimal lepton flavor violation (MLFV) is that there is one, and only one spurion that breaks the global $SU(3)_L\times SU(3)_E$ flavor symmetry, and it transforms as $(3,\bar3)$ under this symmetry.

We emphasize that, while this definition of MFV implies that flavor changing couplings in the quark sector depend on the CKM parameters, the converse is not true: It is not the case that any model where flavor changing couplings are determined by the CKM parameters is MFV. Thus, the models proposed in Ref.
\cite{Botella:2009pq} are not MFV as defined here (and as defined in Ref. \cite{D'Ambrosio:2002ex}). The models of Refs. \cite{Blum:2011fa,Hochberg:2011ru,Nelson:2011us} provide examples of models that are far from being MFV, yet the flavor changing couplings depend on purely CKM parameters.

The implementation of the MFV principle in two Higgs doublet models is less straightforward than it is in a single Higgs doublet model or, more generally, than it is in models with natural flavor conservation. The reason is that, for each of the three spurion representations mentioned above, there are two corresponding Yukawa matrices. We argue that, assuming that there is indeed a single fundamental spurion that breaks, for example, $SU(3)_L\times SU(3)_E$, then one may choose any matrix that transforms in the same way, and whose leading term is linear in the fundamental spurion, without loss of generality.

Several of our results are closely related to those that have been derived in Ref. \cite{Kagan:2009bn} in the framework of `general minimal flavor violation' (GMFV). The GMFV analysis applies to the SM as a low energy effective theory, while our model allows for additional light degrees of freedom. Yet, there are common features due to the fact that for $\hat Y_t,\hat Y_b,\hat Y_\tau={\cal O}(1)$ the respective low energy effective theories can be described by $[U(3)/U(2)\times U(1)]^3$ non-linear $\sigma$ models. In this regard, it is interesting to note that, as far as the entries of $Y^F_h$ are concerned, the various relations that we derive are the same as those derived in the framework of a single Higgs doublet model with the dimension-six terms of the form
$(\Phi^\dagger\Phi)\overline{L}Y^{E\prime}  E \Phi$
(and similarly for the quark sector), where the structure of the $Y^{F\prime}$ matrices is subject to the MFV selection rules \cite{Dery:2013rta}.

We are interested in the phenomenological consequences of MFV for the flavor structure of the scalar couplings in two Higgs doublet models. In particular, we are interested in the question whether measurements of various couplings of the recently discovered boson $h$ within a single fermion sector ($E$, $U$ or $D$) can be used to test MFV. Two Higgs doublet models with minimal flavor violation (defined in a way consistent with ours) were recently studied in  Refs. \cite{Buras:2010mh} and \cite{Altmannshofer:2012ar}. However, the focus of the first is on the implications for flavor changing neutral current processes, and of the latter on the consequences for the LHC of the modifications in couplings to the third generation fermions. Moreover, both use approximations that are appropriate for their study, but cannot be applied in our study.

While we presented all the MFV predictions, not all of the relevant couplings are likely to be accessible in present and near future experiments. The sector that is most likely to be experimentally probed is the charged lepton sector, where $(Y^E_h)_{\tau\tau}$, $(Y^E_h)_{\mu\tau}$, and $(Y^E_h)_{\mu\mu}$ might be measured or constrained. If $h\to\mu^\pm\tau^\mp$ is observed, Eq. (\ref{eq:ysab}) will be violated and MLFV will be excluded \cite{Dery:2013rta}. The ratio $X_{\mu\tau}\equiv{\rm BR}(h\to\mu^+\mu^-)/{\rm BR}(h\to\tau^+\tau^-)$ can be used to test the relation (\ref{eq:univmt}) \cite{Dery:2013rta}. If it is violated, then either MLFV does not apply or $\hat Y^E_\tau$ (and thus $(Y^E_A)_\tau$) is large and, consequently, the correction to (\ref{eq:univmt}) -- see Eq.~(\ref{eq:abmt}) -- is significant. Thus, MLFV models with $(Y^E_A)_\tau\sim1$ provide an example of viable flavor models where the SM relation, $X_{\mu\tau}=(m_\mu/m_\tau)^2$, can be violated by order one corrections.

Another sector where we might hope to have experimental information on more than a single Yukawa coupling is the up sector. Here, $(Y^U_h)_{tt}$ and $(Y^U_h)_{ct}$ can be measured or constrained (see {\it e.g.}~\cite{Craig:2012vj}). There is no exact relation between the two, but we expect $(Y^U_h)_{ct}/(Y^U_h)_{tt}\lsim \hat Y^{2}_bV_{cb}$. Thus, violation of ${\rm BR}(t\to ch)/{\rm BR}(t\to sW)\lsim1$ will be suggestive that MFV does not hold.

\vspace{1cm}
\begin{center}

{\bf Acknowledgements}
\end{center}
We thank Gilad Perez for useful discussions. This project is supported by the German-Israeli foundation for scientific research and development (GIF). YN is the Amos de-Shalit chair of theoretical physics. YN is supported by the Israel Science Foundation.


\end{document}